\documentclass[twocolumn]{article}
\usepackage[utf8]{inputenc}
\usepackage{authblk}
\usepackage{geometry}
\usepackage{layout}

\makeatletter
\let\@fnsymbol\@arabic
\makeatother

\usepackage[dvipsnames]{xcolor}
\usepackage{hyperref}
\usepackage[nameinlink]{cleveref}
\usepackage{graphicx}
\usepackage{enumitem}
\usepackage{color}
\usepackage{microtype}

\graphicspath{{figures/}{pictures/}{images/}{data/}{./}}

\usepackage[
backend=biber,
style=numeric,
sorting=ynt,
maxbibnames=99,
giveninits=true,
style=numeric,
]{biblatex}
\hypersetup{ 
  breaklinks=true,
  colorlinks=true,
  linkcolor={red!50!black},
  citecolor={blue!50!black},
  urlcolor={blue!80!black}
}

\crefname{section}{Sect.}{Sects.}
\Crefname{section}{Section}{Sections}
\crefname{subsection}{Sect.}{Sects.}
\Crefname{subsection}{Section}{Sections}
\crefname{subsubsection}{Sect.}{Sects.}
\Crefname{subsubsection}{Section}{Sections}
\crefname{figure}{Fig.}{Figs.}
\Crefname{figure}{Figure}{Figures}
\crefname{table}{Tab.}{Tabs.}
\Crefname{table}{Table}{Tables}

\newcommand{\rone}{$(\mathrm{R}_1)$}
\newcommand{\rtwo}{$(\mathrm{R}_2)$}
\newcommand{\rthree}{$(\mathrm{R}_3)$}

\SetLabelAlign{myparleft}{\parbox[t]\columnwidth{#1\par\mbox{}}}

\begin{document}

\title{Towards a Unified User Interface for Visual Analysis\\ of Retinal Data in Ophthalmology}

\date{}

\author[,a]{Martin R{\"o}hlig\thanks{martin.roehlig@uni-rostock.de}}
\author[,a]{Lars Nonnemann\thanks{lars.nonnemann@vcric.igd-r.fraunhofer.de}}
\author[,b]{Hans-J{\"o}rg Schulz\thanks{hjschulz@cs.au.dk}}
\author[,c]{\\Oliver Stachs\thanks{oliver.stachs@uni-rostock.de}}
\author[,a]{Heidrun Schumann\thanks{heidrun.schumann@uni-rostock.de}}
\affil[a]{Institute for Visual and Analytic Computing, University of Rostock, Germany}
\affil[b]{Department of Computer Science, Aarhus University, Denmark}
\affil[c]{Department of Ophthalmology, Rostock University Medical Center, Germany}

\maketitle

\begin{abstract}
The visual analysis of retinal data contributes to the understanding of a wide range of eye diseases.
For the evaluation of cross-sectional studies, ophthalmologists rely on workflows and toolsets established in their work environment.
That is, they know what tools and data are needed at each step of their workflow.
Yet, manually operating the various tools, including activation, data handling, or view arrangement, can be cumbersome and time-consuming.
We thus introduce a new visualization-supported toolchaining approach that combines workflow, tools, and data.
First, we provide access to the tools required for each step of the workflow.
Second, we handle the exchange of data between these tools.
Third, we organize the views of the tools on screen using suitable layouts.
Fourth, we visualize the connection between workflow, tools, and data to support the data analysis.
We demonstrate our approach with a use case in ophthalmic research and report on initial feedback from experts.
\end{abstract}
\section{Introduction}\label{sec:introduction}

Over the years, many different visual analytics (VA) tools have been developed to support the analysis of medical data in a wide range of application areas.
The general goal is to help domain experts in their data analyses by providing means for decision making and knowledge discovery.
This is also true for the area of ophthalmology, where VA tools have contributed to the understanding of various eye diseases.
In collaboration with ophthalmologists, we were able to improve the detection of early retinal defects using dedicated VA tools for retinal data~\cite{PRFG19,PMFS20}.
This was made possible by developing the tools specifically for the needs of the ophthalmologists and integrating them into their workflows for the evaluation of cross-sectional studies.

However, a single VA tool is often not enough to fully handle complex analysis workflows.
On the one hand, this is because different VA tools may be needed for different steps of the analysis.
On the other hand, tools for data management and preprocessing are typically also required, as well as for statistical analysis and compilation of final results.
In general, the use of VA tools is focused on the steps necessary for the actual visual data analysis.
The remaining steps are typically performed with other software solutions, ranging from clinical information systems over examination device-specific software to general spreadsheet and statistics software.
The reason for this is that trying to integrate all the functionality into a single tool would significantly increase the effort required for software development and maintenance.
In certain cases it may even simply not be possible due to time and resource constraints or because of the proprietary algorithms or data formats used.
As a result, domain experts have to interact with various tools -- VA-based or not -- and exchange data between them to fulfill all the necessary steps in their analysis workflows.

One of the resulting problems is that for a given analysis workflow in the domain, it may not be immediately clear how to coordinate the required tools and the corresponding visual output on the screen.
For each workflow step, domain experts must determine \emph{which} tools and \emph{what} pieces of information to show before deciding \emph{how} to display them on the screen to get the current step done.
This places an additional burden on a given workflow, making it more cumbersome and time-consuming to execute.
At the same time, no support is generally offered to complete these secondary tasks alongside the actual data analysis.
Instead, in practice, experts have to manually switch between different stages of preprocessing, visual data analysis, and result interpretation, and operate the tools and their output by hand.
Often, this is necessary again and again each time a workflow is executed.

We present a novel visualization-supported toolchaining approach that combines workflow, tools, and data.
With a use case in the analysis of retinal data in ophthalmology, we show that our approach not only provides access to the tools and data needed for each workflow step, but also helps to organize the tools' user interfaces (UIs) on the screen.
This reduces the overhead of managing the various tools and data during the workflow execution.
In our collaboration with ophthalmologists, the experts were thus able to concentrate on the actual steps of data analysis in the evaluation of cross-sectional studies.
Our contributions are:

\begin{description}
\item[Integration of workflow and tools:]
We examine existing connections between workflow, tools, and data in the context of our use case and establish meaningful links for accessing the right tools, the right data, and the right views at the right time.

\item[Visualization support for toolchaining:]
We present an editor that allows to interactively create toolchains and adjust tools and data for each step of a workflow.
We incorporate a unified UI that shows the workflow, tools, and data and helps to organize multiple tool views on screen.

\item[Application in practice:]
We assess our approach together with domain experts in a use case in ophthalmology.
We discuss the differences with current data analysis practices and report initial expert feedback.
\end{description}

The remainder of this paper is structured as follows.
The background and current analysis workflow are described in \Cref{sec:background}, followed by an overview of related work in \Cref{sec:related}.
Our approach to combining workflows, tools, and data is detailed in \Cref{sec:approach}.
An application example and user feedback are presented in \Cref{sec:application}.
Finally, a conclusion and topics for future work are discussed in \Cref{sec:conclusion}.
\section{Background}\label{sec:background}

In ophthalmology, data from various examination methods are analyzed to understand the impact of different eye diseases on the condition of the human retina.
One widely used examination method is optical coherence tomography (OCT)~\cite{HSLS91,HISH95}.
Modern OCT scanners enable noninvasive imaging of substructures of the multilayered retina with high spatial resolution~\cite{DMGK01}.
By analyzing the OCT images, ophthalmologists are able to detect common ocular diseases, such
as age-related macular degeneration~\cite{KPLH12}, diabetic retinopathy~\cite{VMCH15}, or glaucoma~\cite{GrTa13}, as well as other pathologies with ocular signs, such as multiple sclerosis~\cite{DWBG11}.
OCT examinations are therefore now a standard procedure in clinics and an integral part of ophthalmic research~\cite{FuSw16}.

We are particularly interested in OCT data analysis in the context of cross-sectional studies in ophthalmic research.
Here, multiple OCT datasets have to be analyzed together with information from other clinical records to compare the retinal condition of patients with that of healthy controls.
This is a complicated process.
To handle the study data, ophthalmologists rely on workflows and toolsets that are established in their work environment.
These workflows can be divided into three common stages:

\begin{description}
\item[Data preparation:]
The required data from all study participants are collected and processed, including filtering and screening based on the study criteria.
Afterwards, the patient and control groups are assembled.

\item[Data analysis:]
With the prepared data, the differences between the patients and controls are computed.
Data subsets of interest are selected, measured, and statistically quantified.

\item[Summarization of results:]
The findings most relevant to the study design are compiled.
The final results are presented by a combination of images, plots, and tables.
\end{description}

Throughout these three stages, ophthalmologists are dealing with different tools for different types of data.
The tools used for the first stage typically include clinical information systems, OCT device-specific software, and general spreadsheet software for managing and processing the raw electronic health records and OCT datasets. 
In the second stage, mainly statistical software is used to analyze the prepared data and, more recently, also exploration-oriented VA tools~\cite{RSPS18,RPSS19}.
Here, the data are often enriched with computed measurements as well as reduced to relevant subsets.
Finally, in the last stage, selected results are summarized and presented with charting tools.

This mixture of different tools, ranging from general-purpose to visualization-specific, has to be operated manually in current analysis practice.
Switching between the tools can be difficult, however, as ophthalmologists must remember which tools are used in which part of the workflow, locate those tools at the right time, and activate them by hand.
The exchange of data can also be quite labor intensive due to the different data formats used and the limited compatibility between the various tools.
It is therefore not easy to keep track of which data was transferred when and between which tools.
On top of that, the output of the tools must be arranged sensibly on the screen, since for some workflow steps not only one, but several tools must be operated simultaneously.
This is the case, for example, in the data exploration stage, where it is often necessary to go back and forth between selection of data subsets, their visual analysis, and statistical quantification.
All in all, the limited integration of workflow steps, tools, and data commonly increased the time and effort of performing an analysis workflow and places an additional burden on ophthalmologists.

Our objective therefore is to help ophthalmologists to show the right tool at the right time, to illustrate the flow of data between them, and to make clear which output was generated for which step with which data.
Establishing this level of support will help reduce the overhead of managing the various tools and data during workflow execution.
\section{Related work}\label{sec:related}

The development of workflow-based toolchains relates to challenges regarding sustainable data exchange~\cite{NSUA20} and tool interoperability~\cite{GAEL16}.
While these two research areas are complex in their own right, dealing with multiple VA tools introduces an additional problem for workflow coordination, namely unifying the UIs of independent tools.

There are many examples for managing multiple views, reaching from comprehensive application toolkits with built-in VA tools to loosely coupled, concurrently running applications~\cite{DCCW08,Robe07}.
Most of the existing solutions rely on the following approaches:

\begin{description}
\item [Integrating views:]
Applications like Tableau~\cite{Stolte.2002}, Dashiki~\cite{McKeon.2009}, or Fusion~\cite{NCIS03} allow users to rapidly assemble multiple views within a given interface.
This is usually achieved via web interfaces to create mashups~\cite{PND10} or webcharts~\cite{FDFR10}.
If enough screen space is available, even dozens of individual views can be incorporated in the same environment~\cite{Langner.2018}.
However, using integrative approaches also has its limitations, as the visual load increases with the number of views included.
    
\item [Coupling views:]
Instead of nested or tabbed UIs, some approaches rely on loose coupling of independent views for view orchestration.
This way, the views are interactively assembled into a common interface.
Example are WinCuts~\cite{TMC04} and Fa\c{c}ades~\cite{SCPR06}, where arbitrary view regions are replicated in a combined interface to focus on task-related areas.
Other methods rely on implicit or explicit visual links between independent views to facilitate the connection between them~\cite{Waldner.2010,Fourney.2014}.
\end{description}

Both the integration of views and the coupling of views depend on the number of views and the available screen space.
This is where the workflow-based order of tool execution becomes important.
By providing a toolchain, information is made available about which tools need to be active at what time and what data needs to be passed from one tool to the next.
This can reduce the visual load for the current analysis step.
While this can help users with the data analysis, it also means that management of tool views over time must be considered.
In this regard, two basic types of displaying tool views can be distinguished:

\begin{description}
\item [Sequential display:]
In simple workflow scenarios, it can be sufficient to display information via only one tool view at a time.
An example is Stack’n’flip \cite{Streit.2012}, where views are displayed in a sequential order and can be switched interactively through a central interface.
    
\item [Parallel display:]
For certain analysis tasks of more complex workflows, it is necessary to execute more than one tool at a time, with the individual views displayed concurrently.
One example is ManyVis~\cite{RSDB13}, where multiple views are shown together in an integrated visualization environment.
\end{description}

When coordinating multiple tools, the toolchain itself is also sometimes encoded visually, e.g., as a directed graph showing its connections~\cite{TIC09,GAEL16}.
Considering the steps completed and data generated with the tools over time, similar visualization approaches can be found in the field of provenance visualization, especially with regard to workflow provenance and data provenance~\cite{HDB17}.
Examples are AVOCADO~\cite{SLSG16} for the visualization of workflow-derived biomedical data, and the systems VisTrails~\cite{BCCF05}, Galaxy~\cite{GNT10}, or Taverna~\cite{WHFW13} for automated tracking and visualization of workflow and data provenance.
The representations of workflow steps and related data with these systems help users understand the data analysis and make it reproducible.
However, combined support for the execution of workflow steps, the sequential or parallel use of tools and their data exchange, and the arrangement of visual outputs on the screen have hardly been addressed so far.

Against this background, we recently introduced a layered toolchaining approach for VA~\cite{SRNH20}.
Our main idea is to model the VA tools in a given workflow as a graph that represents the tools as its nodes and the different levels of communication between them as directed edges.
This provides flexibility for describing different couplings, regardless of whether the tools are used in a sequence, repeatedly back-and-forth, or simultaneously side-by-side.
Based on this model, we were able to characterize the data exchange between VA tools~\cite{NSUA20} and introduce ReVize~\cite{HoSc19}, a library for adding toolchain support for web-based applications using Vega-Lite~\cite{SMWH17} as a common exchange format.
We further demonstrated a framework to interactively create toolchains via custom data connections between independent VA tools~\cite{NHSU21,NHRS22}.
We have not yet addressed, however, how to present and interact with a unified UI and how to incorporate the tools' output into a sensible assembly during the execution of a workflow in practice.

In summary, there are several approaches for the unification of UI elements.
With respect to retinal data analysis in ophthalmology, it is not clear how to translate these approaches into practice, especially when considering complex analysis scenarios such as the evaluation of cross-sectional studies.
We therefore aim to fill in the missing pieces and develop a consistent coordination approach for our application example.
To this end, we build on our ideas of layered toolchaining~\cite{SRNH20} and investigate (1) what tools to show in which part of the workflow, (2) what data parts to exchange between the tools, (3) how to present the tools and the data on screen, and (4) how to establish a connection between workflow, tools, and data that supports the analysis of retinal data.
\section{Workflow-based toolchaining for visual analytics}\label{sec:approach}

We aim to support toolchaining for VA of retinal data in ophthalmology.
To find suitable solutions, we first consider the requirements in our use case.
On that basis, we then present new concepts and visual designs.

\subsection{Requirements}

We conducted interviews with ophthalmologists and job observations to get an overall idea of the current analysis practices.
Together, we then engaged in the statistical analysis of two studies~\cite{GKPJ18,PGKJ18} and subsequently two additional studies~\cite{PRFG19,PMFS20} analyzed with a mixture of statistical and VA tools.
This allowed us to gain deeper insights into their workflows and use of VA, statistics, and general-purpose analysis software.
We finally compiled a list of design requirements related to the coordination of VA tools for retinal data analysis.

\begin{description}
\item[Access to tools \rone{}:]
In a workflow, the temporal order of tools and their interplay is typically only implicitly defined.
On the one hand, this means that the ophthalmologists must remember which tools to activate for the current work step.
On the other hand, they also need to determine what content to display at the same time with these tools.
The contents include currently required panels and views, but also information about available data and information about already used or subsequent tools.
Therefore, access to the right tool at the right time must be supported.

\item[Connection of tools and data \rtwo{}:]
In general, the entire data are not passed between analysis tools over and over again during the execution of a workflow.
Instead, the data are incrementally reduced and sometimes even selectively extended to extract valuable information.
Thus, ophthalmologists must not only supply the original data to the tools at the beginning of the workflow, but also manage the exchange of various data pieces between the tools during subsequent steps.
This includes taking care of all necessary data transformations between the tools and updates if parts of the data have changed during the analysis.
Supporting access to the right data at the right time and establishing a link between the data and the tools is thus a requirement.

\item[Arrangement of tool views \rthree{}:]
When activating different tools, the layout of their views should correspond to their intended use in the respective part of the workflow.
If the tools are used one after the other, their views may simply be interchanged.
However, if they are used in parallel, their views must be organized on the screen in a meaningful way so that they can be used together effectively.
In addition, navigational features are needed that not only give the user control over switching between workflow steps and associated view layouts, but also help to maintain orientation at the same time.
Consequently, support for obtaining the right view layout at the right time is needed to manage the visual output of tools.
\end{description}

From a VA perspective, the three requirements \rone{} to \rthree{} are related to three fundamental questions that need to be answered in order to support toolchaining in our use case.
The first questions is \emph{what to show?}.
With respect to requirement \rone{}, it addresses the selection of all relevant information that needs to be presented to the user at a given point in time.
The second question is \emph{what data parts to exchange?} and corresponds to requirement \rtwo{}.
It aims to identify subsets of interesting data and consider when and how they should be exchanged between tools.
The third question is \emph{how to show?}.
Related to requirement \rthree{}, it is primarily about the layout and adaptation of views and panels of the current tools.
In addition, it includes presenting information about the workflow itself and how workflow steps, data, and tools are connected.
By answering these three questions in our VA design, we create a meaningful coordination between workflow, tools, data, and view layouts.

\subsection{Combination of tools, data, and view layout}

\begin{figure}
\centering
\includegraphics[width=\columnwidth]{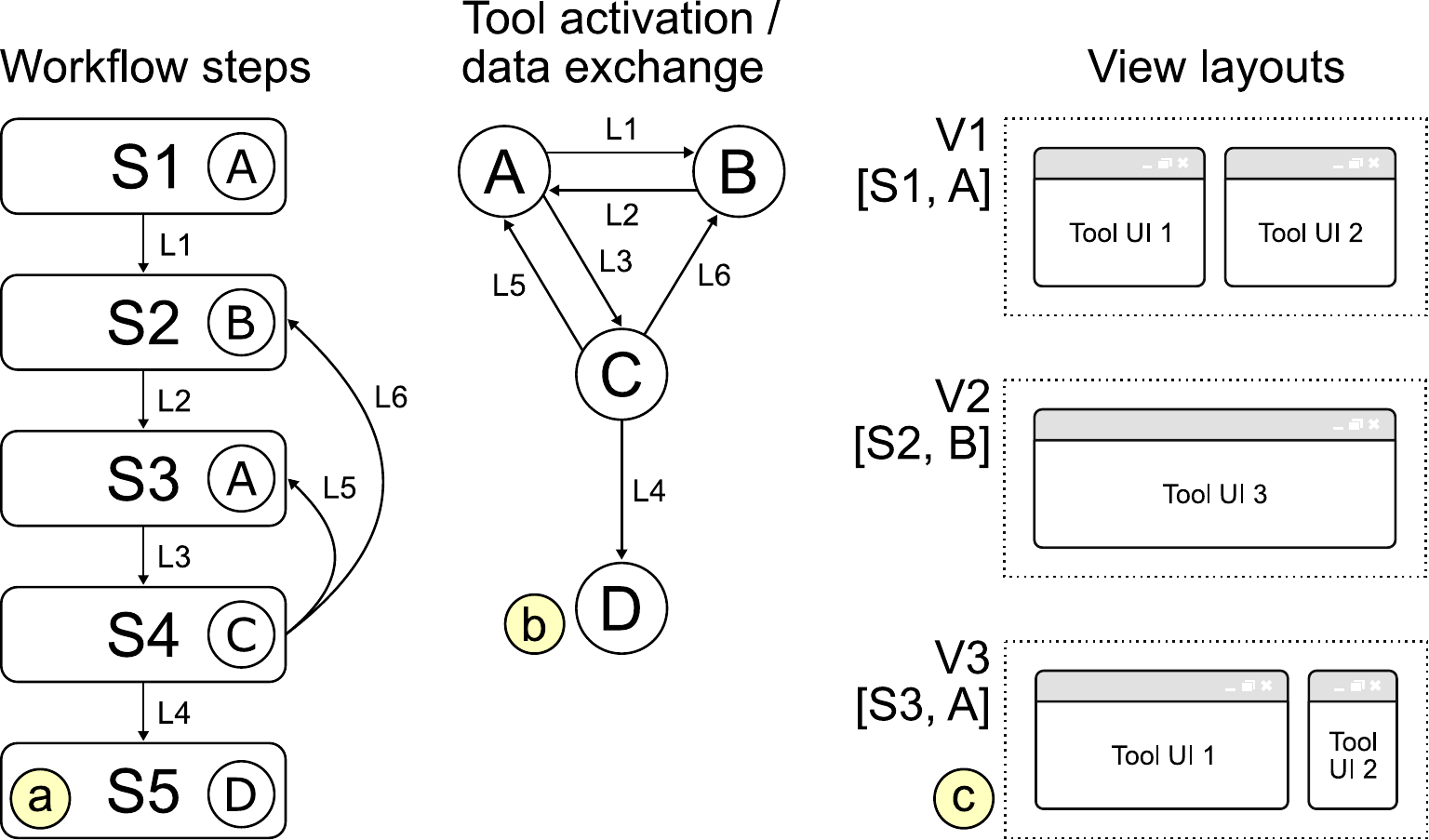}
\caption{Overview of the coordination components.
Shown are the workflow steps \emph{S1} to \emph{S5}~(a), associated toolsets \emph{A} to \emph{D}~(b), links between workflow steps and tools \emph{L1} to \emph{L6}, and view layouts \emph{V1} to \emph{V3} assigned to each workflow steps and toolset~(c).}
\label{fig:approach:components}
\end{figure}

To meet our design requirements and answer the three corresponding questions, we first need to determine several pieces of information.
For this purpose, we build on our earlier work on a layered approach to lightweight toolchaining in VA~\cite{SRNH20}.
In particular, we implement and adapt the proposed coordination model to establish a connection between the four components: (i) domain workflow, (ii) tools used, (iii) data analyzed, and (iv) layout of tool views.
\Cref{fig:approach:components} illustrates the interplay of these components.

We start by looking at the existing workflow and the tools used in the application domain.
This allows us to determine a set of tools per workflow step (\cref{fig:approach:components}a).
Generally, one or more tools may be required to perform each step.
The sequences in which these tools must be activated can be derived from the order of the workflow steps.
The back and forth between the steps also determines which data connections must be established between which tools.
With this information at hand, we build up a coordination graph~\cite{SRNH20} by stepwise defining a link between pairs of tools (\cref{fig:approach:components}b).
The links capture both the usage flow, i.e., all possible activation sequences, and the data flow, i.e., all possible data exchanges, between the tools.
Based on this, we then define different arrangements of tool views and assign a suitable layout to each combination of workflow step and toolset (\cref{fig:approach:components}c). 
Note that one set of tools can be used in multiple workflow steps, but the content of the tools can be displayed with a different layout specific to each particular step.
We determine these layouts based on user preferences.
For example, we support choosing from a set of default layouts or saving custom arrangements of views per step.

The final coordination graph provides all the information necessary to facilitate toolchaining in our use case.
That is, it allows us to query at any time which tools need to be activated or deactivated, which pieces of data need to be transferred from a current tool to the next, and which content should be displayed how on the screen.
In this way, it enables us to answer in particularly the questions \emph{what to show? \rone{}} and \emph{what data parts to exchange? \rtwo{}}.
Incorporating different view layouts into the graph further provides the basis for answering the third question \emph{how to show? \rthree{}}.
The actual path through the coordination graph, however, is not determined until the workflow is executed at runtime and the user interactively decides which steps are to be processed in which order.
Hence, in addition to the arrangement of the individual tools on the screen, it must be possible to access the coordination graph, the progress of the workflow, and the results obtained during its execution.
To this end, we present a new design that provides this information both visually and interactively, specifically addressing the third question \emph{how to show? \rthree{}} again.

\subsection{Visualization support for the unification of UI ensembles}

The workflow describes the analysis steps to be executed by the user.
Per step, one or more tools are used.
The coordination graph captures the pairwise coupling of the tools and the associated data exchange.
The assigned layouts determine how the tool views are displayed on the screen.
This results in a toolchain that controls the handling of the tools, including going back and forth between them.

To use the coordination graph effectively, the information it contains must be made available to the user.
We therefore introduce a unified UI that visualizes the workflow together with the coordination graph.
Our design assists users in creating or adapting a coordination graph, retrieving information from it for the execution of the workflow, and comprehending the results obtained.
It consists of several views that allow switching between the current work step and the corresponding toolchain in the background.

\paragraph*{Visualizing the workflow and coordination graph:}

\begin{figure*}[ht]
\centering
\includegraphics[width=\textwidth]{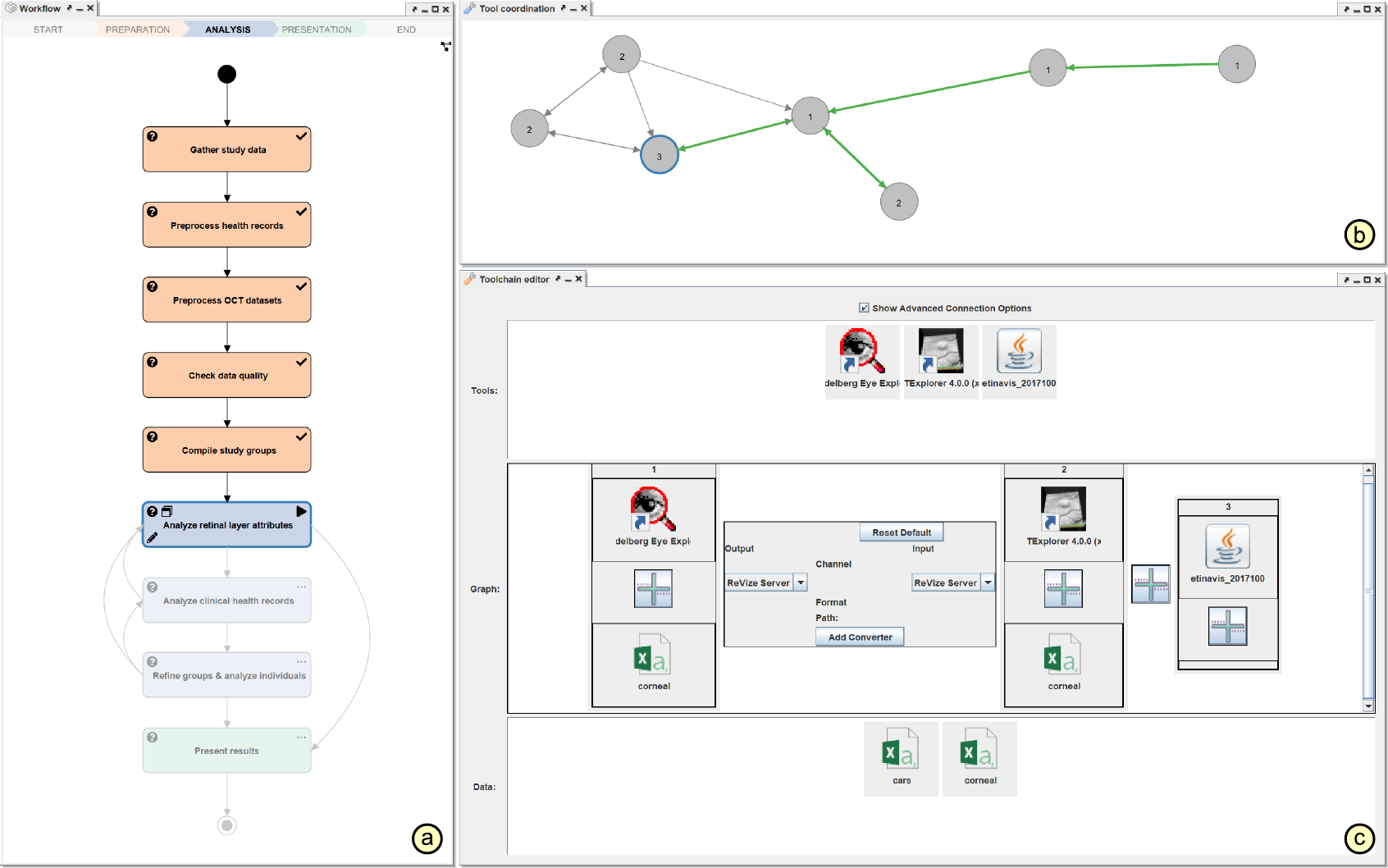}
\caption{Visualization of workflow, coordination graph, and toolchain editor.
The workflow steps are displayed as colored rectangles~(a).
The coordination graph is represented as a network visualization, where the circles depict the tools and the lines encode the links between them~(b).
The toolchain editor~(c) consists of three panels showing the available tools (top), the data sources (bottom), and the data connections between tools (middle).}
\label{fig:approach:workflow_coordination}
\end{figure*}

From the user's point of view, it is important to get an idea of the work at hand before executing a particular workflow.
This involves not only the individual workflow steps that must be performed, but also the data to be analyzed and the extent to which tool coordination is supported.
Therefore, with our unified UI, we provide an overview of the workflow along with the coordination graph.
A toolchain editor makes it possible to view individual links between tools in detail and make adjustments as needed.
\Cref{fig:approach:workflow_coordination} illustrates our design consisting of three views.

The \emph{workflow view} displays the workflow steps as a flowchart (\cref{fig:approach:workflow_coordination}a).
The steps are shown as rectangles colored according to their workflow stage (preparation, analysis, summarization).
The arrows between the rectangles encode the possible back and forth between the steps.
Information about the current progress within the workflow is provided by highlighting the steps and links that have already been performed.

The \emph{coordination graph view} shows the toolsets used in the workflow as a network visualization (\cref{fig:approach:workflow_coordination}b).
The toolsets are coded as circles and the lines indicate their data exchange and activation capabilities.
Hovering over the circles displays additional information, e.g., which tools are included in the respective set and in which steps they are used.

The third view consists of a \emph{toolchain editor} (\cref{fig:approach:workflow_coordination}c).
Based on our previous work~\cite{NHSU21,NHRS22}, the main purpose of the editor is to establish a connection between the workflow and the coordination graph.
It consists of three panels showing all available tools (top), all data sources (bottom), and the links between the tools (middle).
Within our interface, the editor performs two functions.
The first is to create a coordination graph, if not already available, for a given workflow and set of tools.
Using the three panels, links between tools can be added step by step via drag \& drop.
The second function is to specify the data sources to be used as input for a planned execution of the workflow, and to adjust the data transformations between tools.
Together, the three views provide an overview of the work, the tools, and the data involved, and allow tool coordination to be set up as required.

\paragraph*{Supporting the workflow execution:}

\begin{figure*}[ht]
\centering
\includegraphics[width=\textwidth]{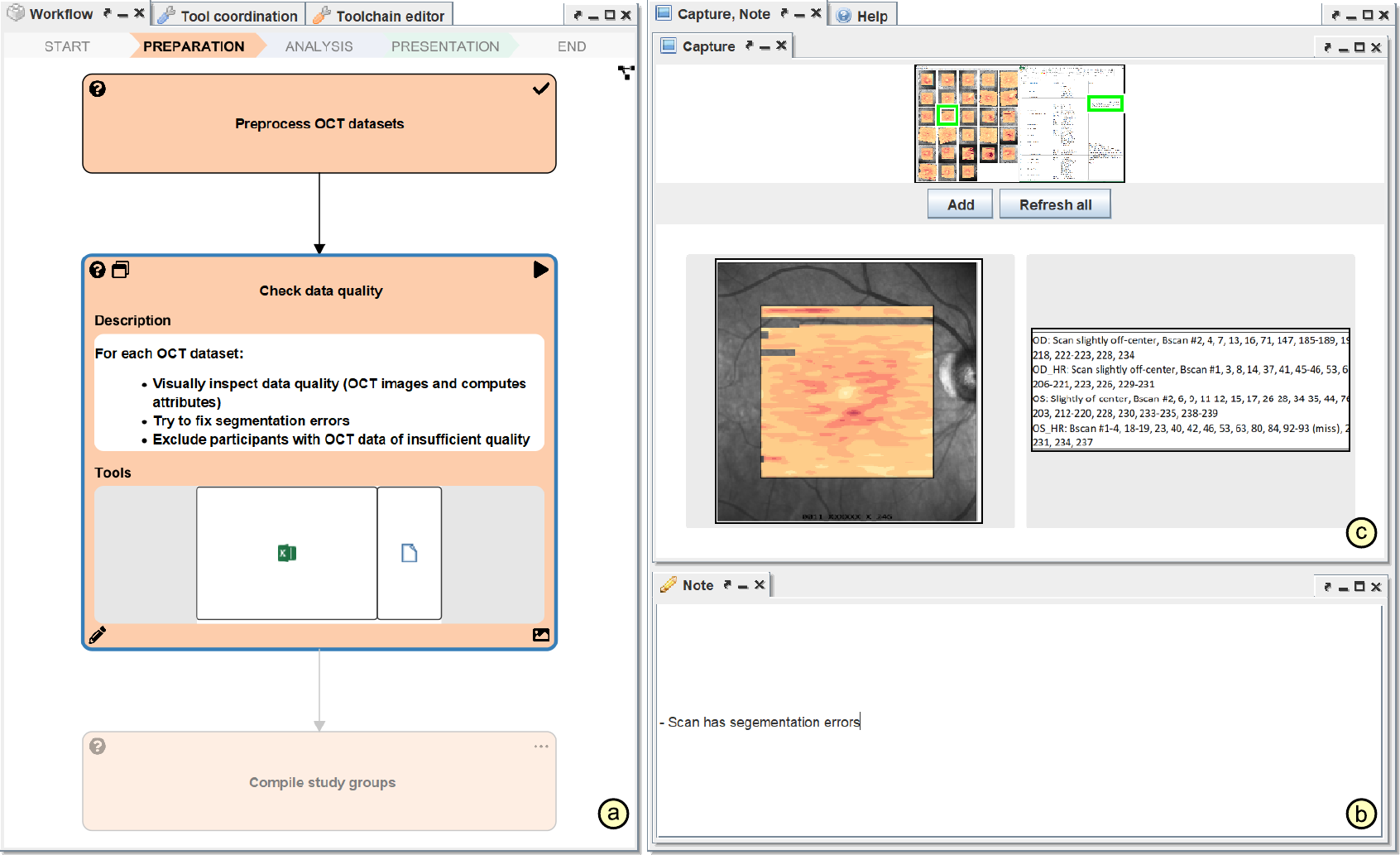}
\caption{Visualization of individual workflow steps and additional information.
On the left side, the current step is enlarged to show additional information, and the previous and next possible steps are displayed above and below~(a).
On the right side, a full description and user notes about the current step are depicted~(b).
In addition, intermediate analysis results are displayed as a list of screen captures~(c).}
\label{fig:approach:step_result}
\end{figure*}

After a connection between the workflow and the tool coordination through the coordination graph has been established, the execution of the individual workflow steps must be supported.
Here, it is important to inform the user about what currently needs to be done, as well as provide navigation support in terms of what has been done before and what comes next.
In addition, the ability to document the work and progress is required.
As illustrated in \Cref{fig:approach:step_result}, our interface support this by showing current steps in detail along with additional information about the coordination.

To focus on the current steps during a workflow execution, the workflow view is switched from an overview to a detail view (\cref{fig:approach:step_result}a).
This enlarges the active step and shows a description of what needs to be done and how the tools needed to do it are arranged on the screen.
The previous step is displayed above the active step and the next possible steps are displayed below it.
Clicking on these steps allows to go back and forth within the workflow.
Once a workflow step has been selected, the corresponding tools are automatically activated and their content is arranged on the screen based on the assigned layouts.
The data exchange between the last and the current tools is also handled automatically as defined in the coordination graph.
This reduces the effort of manually searching for the right tools, activating them, transferring the work data, and customizing where the content is displayed on screen.
To see the actual path taken so far through the workflow and coordination graph, the user can switch back to the overview visualizations (\cref{fig:approach:workflow_coordination}) at any time.
This highlights the links of the active step and tools in the overviews, making adjustments in the coordination graph possible on the fly.

Our interface also helps to document the work performed.
On the one hand, the user can set the status of the active workflow step (e.g., pending, done, paused, canceled) to indicate the state of the work.
In addition, notes can be added to the steps describing how the work was performed and whether it was necessary to deviate from the original work description (\cref{fig:approach:step_result}b).
On the other hand, intermediate results can be captured via screenshots of the tool content currently visible on the screen (\cref{fig:approach:step_result}c).
Multiple screenshots can be added by selecting any areas of the screen, with a reference image showing which areas have already been captured.
The content of the selected screen areas can be updated or removed again if relevant changes occur during work.
If a workflow step is visited multiple times, e.g., by navigating back and forth in the workflow, a new set of notes and captures is created for each activation.
That way, a detailed history of all work performed is generated.
Especially when different paths are taken through the workflow, this helps to recapitulate which steps were taken to arrive at the results achieved in each case.

\paragraph*{Compiling the results:}

\begin{figure*}[t]
\centering
\includegraphics[width=\textwidth]{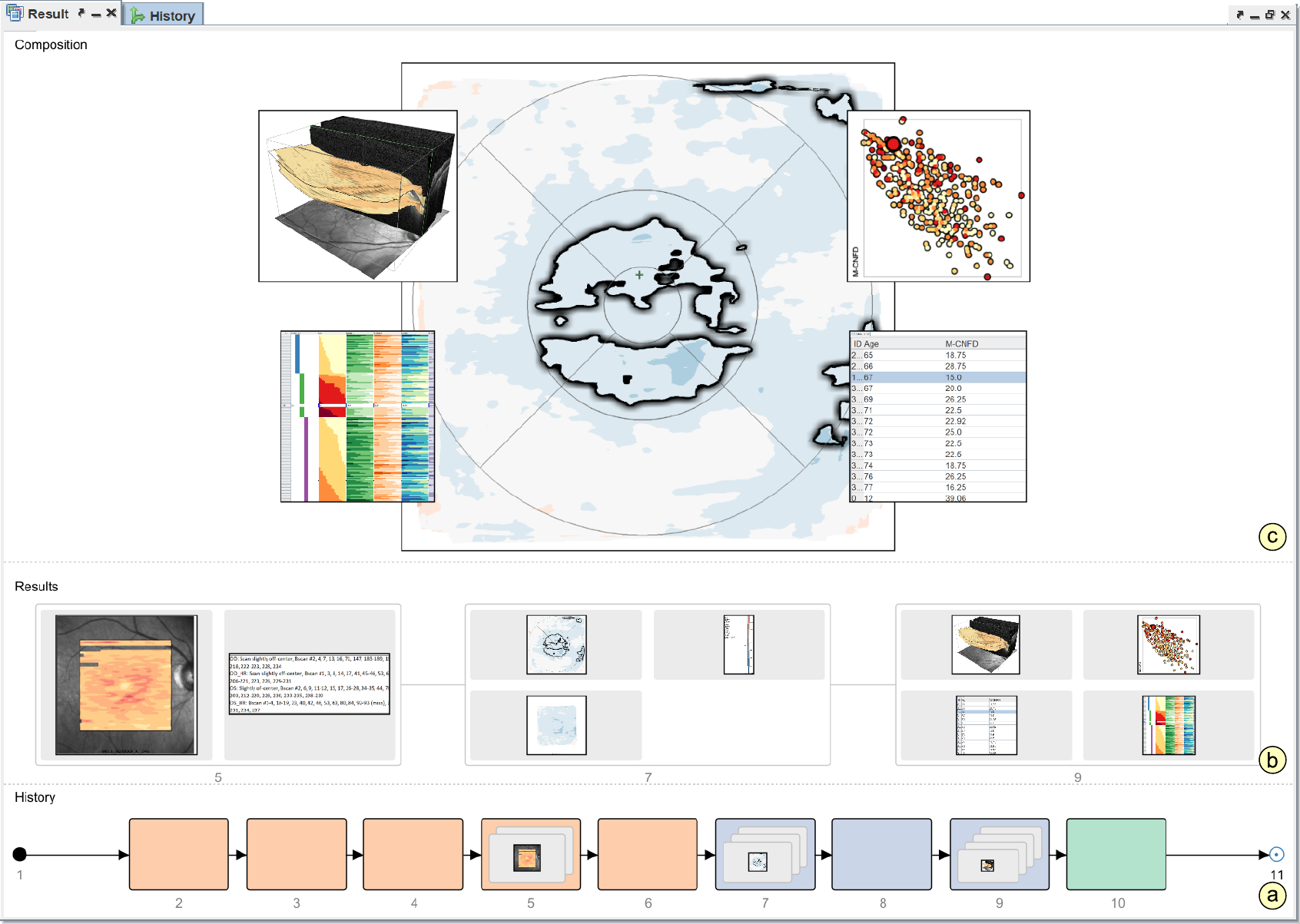}
\caption{Visualization of workflow history and result composition.
At the bottom, a history of all performed workflow steps is displayed~(a).
In the middle, all captured intermediate results are listed~(b).
At the top, selected intermediate results are summarized in an overview that illustrates the main findings of the data analysis~(c).}
\label{fig:approach:history_composition}
\end{figure*}

During and after the execution of the workflow, it typically becomes necessary to summarize the work performed.
Going back and forth between workflow steps and multiple tools, however, can make it difficult to recapitulate the work after completion and extract the most meaningful insights from multiple intermediate results.
In fact, the end results do not have to come from a single workflow step and a single set of tools.
It is rather often a mixture of intermediate results from multiple steps and tools that need to be reconciled and compared to provide an overall picture of the work done and insights gained.
Therefore, our interface provides an additional \emph{summary view} for compiling results across workflow steps and tools.
\Cref{fig:approach:history_composition} illustrates this view.

The summary view consists of three areas.
In the lower area, a history of all performed workflow steps is displayed in the order of their activation (\cref{fig:approach:history_composition}a).
Each step is represented by a colored rectangle with small icons indicating whether intermediate results have been captured.
In the middle area, the intermediate results are listed grouped by the respective steps (\cref{fig:approach:history_composition}b).
In the top area, the intermediate results can be interactively combined into a single image via drag \& drop (\cref{fig:approach:history_composition}c).
The finished image can then be exported for further use.
This allows to bring together and compare the main findings of the data analysis.
Without our unified UI, this would hardly be directly possible if only the work steps and the associated tools were considered individually.

Overall, the different views of our unified UI establish a visual connection between workflow, tools, data, and displayed content.
Our design approach thus assists users in creating a coordination graph that fits a particular workflow and provides valuable information for executing the workflow and understanding the results.
To assess the utility, we tested our approach with a use case in ophthalmology.
\section{Application to retinal data analysis}\label{sec:application}

We applied and tested our solution in collaboration with ophthalmologists.
As briefly described in \Cref{sec:background}, the ophthalmologists were particularly interested in the analysis of retinal OCT data in the context of cross-sectional studies.
In such studies, data from a group of patients are compared with data from a control group to determine any influence of the disease under study on the condition of the retina and other measured clinical parameters.
After the data acquisition from all study participants, the study evaluation follows established workflows that are usually divided into three stages: (1) data preparation, (2) data analysis, and (3) summary of results.
Currently, however, the ophthalmologist must manually operate the tools required to accomplish these stages.
Together, we therefore created a coordination graph according to their steps and tools, and compared the execution of the workflow supported by our solutions with their experience with manual tool coordination in previous studies~\cite{GKPJ18,PMFS20}.

\subsection{Connection of workflow, tools, and data}

For our use case, we reevaluated the data of a previous study~\cite{PMFS20} with focus on the detection of thickness changes in intraretinal layers at an early stage of diabetes mellitus.
In this study, data from 33 diabetic patients and 40 healthy controls were analyzed.

\paragraph{The data:}
In the study evaluation, we distinguished between two types of input data.
The first type was retinal OCT datasets.
These datasets consisted of 3D images and segmented layer boundaries of each participant's retina.
The second type was clinical health records.
The record of each participant included various parameters such as age, body mass index, blood glucose level, and type of medication received.

\paragraph{The tools:}
A total of 8 different tools were used to process and analyze the data.
On the one hand, these were commercial software tools specifically for the acquisition and processing of retinal OCT data.
For the management of the tabular data from the electronic health records also general spreadsheet and statistical software was used.
On the other hand, we applied our own VA tools designed for the exploration of retinal OCT data and multivariate data from clinical health records~\cite{LRKS14,RSPS18,RPSS19}
For statistical analysis, we additionally used custom scripts together with the R software environment~\cite{RLan21}.

\paragraph{The workflow:}
For the evaluation of the study data, we built on our prior research on workflow-based visual analysis of retinal data~\cite{RPSS19}.
Together with the ophthalmologists, we enhanced our approach to include the multiple tools and data sources needed for comprehensive ophthalmic research.
In the extended workflow, the ophthalmologists had to perform 9 steps (\cref{fig:approach:workflow_coordination}a).
In the first 5 steps, they focused on the data preparation.
This included collecting study data from all subjects, calculating retinal thickness, checking data quality, and assembling the two groups to be compared.
In the next 3 steps, the ophthalmologists continued with the data analysis.
Here, their main focus was on analyzing the differences in thickness values between the groups.
They also alternated between the analysis of OCT data and clinical parameters to identify outliers and refine the groups accordingly.
For this purpose, a mixture of visual exploration and statistical hypothesis testing was used.
Especially at this stage, there was therefore a back-and-forth between workflow steps to narrow down the data to relevant subsets.
In the final step, the ophthalmologists summarized the key findings based on their analysis of the identified subsets.
This included extracting images of relevant data values and creating additional charts, e.g., to highlight interesting statistical results.

\paragraph{The toolchain:}
The tool coordination for our use case was created with the toolchain editor (\cref{fig:approach:workflow_coordination}c).
First, we imported all tools and datasets into the editor.
Together with the ophthalmologists, we then focused on making the tools available when needed by automating their activation.
For this purpose, we displayed the workflow steps in the overview visualization and created corresponding links between the tools using the editor.
The visual representation of the connections in the editor helped us to get an idea of when which tool was used. 
Using the drag \& drop features of the editor, we then set the data input for the workflow and customized the data transfer between the tools.
Once the links were established, we performed a test run by navigating the workflow to ensure that the modeled coordination graph met the ophthalmologists' expectations.
The test run also served to assign a suitable view layout to the automatically activated tools for each workflow step.
Based on the default layouts, the ophthalmologists were able to refine where content should be displayed and how much space each view was given on the screen.
The layouts, along with a description of the workflow and coordination graph, were saved for subsequent executions.

\subsection{Workflow execution and testing}

We applied the workflow and toolchain according to the use case defined together with ophthalmologists.
During conception of our approach and the testing of our solutions, we obtained initial informal feedback in discussions with primarily two ophthalmic experts.

\paragraph{Execution of the workflow and toolchain:}

After the initial configuration described above, we executed the previously saved descriptions of the workflow and toolchain.
Starting from the overview of the workflow (\autoref{fig:approach:workflow_coordination}a), the UI was switched to the detail view (\autoref{fig:approach:step_result}a) to walk through the individual steps one by one.
At each step, the assigned tools were activated and their views automatically arranged on screen.
Based on the displayed description of the actions to be performed, the tools were operated with the supplied data.
Where necessary, we were able to review the underlying data exchange and transformation in the toolchain for the current step by switching between the workflow view and the toolchain editor (\autoref{fig:approach:workflow_coordination}c).
Throughout the workflow execution, user comments and screenshots of intermediate results were collected using the developed UI controls (\autoref{fig:approach:step_result}b, c).
These results were then summarized in a final overview of the main findings, while the visual workflow history helped to recapitulate in which stage and step each part was obtained (\autoref{fig:approach:history_composition}).

Overall, the same medical findings were extracted and reproduced with our visualization-supported toolchaining approach that we gained in our earlier analyses of the same study data~\cite{RPSS19,PMFS20}.
This time, however, we were able to support all three stages of the workflow with our unified UI, from data preparation over data analysis to summary of results.
Compared to manual coordination in previous studies, this reduced the effort required to activate the various tools, manage the exchange of data between them, and orchestrate their views for sequential and parallel display.

\paragraph{Discussion and user feedback:}

Given the results of the workflow execution, the feedback from the experts was largely positive.
In particular, they pointed to the benefits of having a visual representation of the workflow along with access to related tools and data through the unified UI.
Likewise, support for different layouts for arranging the tool views for each work step on the screen was considered advantageous.
They also liked the idea of summarizing their results based on the intermediate comments and screenshots directly in the UI, but noted that a future version of the configured toolchain may need to include more tools to produce final figures and tables for reporting the study results.
Nevertheless, they appreciated our solutions towards a unified UI for retinal data analysis.

On the other hand, given the heterogeneity of the tools, not all steps of the use case could be fully automated during workflow and toolchain configuration.
Further fine-tuning of our general solutions could help to accommodate some of the specific steps involved in the evaluation of this type of ophthalmic study data.
In general, however, we agreed to coordinate only what is necessary, as opposed to what is theoretically possible, to balance the effort of configuring the workflow and toolchain and developing additional controls in the unified UI.
Related to this, we discussed the work that must be invested up front to force the coupling of otherwise seemingly incompatible tools, given the amount of actual support gained during execution and the generalizability beyond a particular use case.
In the end, our discussions led to several directions for future improvements (\cref{sec:conclusion}).

Finally, we noted the need for more in-depth testing and formal evaluation of our methods to fully assess the benefits and limitations of our solutions beyond the preliminary results described here.
Therefore, we are currently continuing our research together with ophthalmologists to further develop a unified UI for retinal data analysis.
Specifically, we will also consider multi-tool analysis workflows with other data types and study methods, including retinal changes associated with cystic fibrosis and longitudinal treatment effects in breast cancer patients~\cite{SBSG22}.
\section{Summary and future work}\label{sec:conclusion}

We presented a new toolchaining approach for VA of retinal data in ophthalmology.
Our approach consists of two parts.
The first part is a coordination graph that captures the interplay of workflow steps, tools used, data to be analyzed, and tool contents to be displayed on the screen.
The second part is a unified UI, which allows to create and customize such a graph for a given workflow, to access all the information needed to execute the workflow, and to understand the results obtained.
By bringing the two parts together, we were able to not only answer the three fundamental questions \emph{what to show?}, \emph{what data parts to exchange?}, and \emph{how to show?}, but also meet the requirements of our collaborating ophthalmologists.
The initial user feedback indicates that our solutions are useful for evaluating cross-sectional studies and reduce the coordination overhead compared to the current manual synchronization of individual analysis tools.

Regarding future work, we see several directions to further support the integration of workflows and toolchaining.
For example, we currently display the contents of the tools on the screen by fixed layouts of the individual tool windows per workflow step.
This prevents the user from having to manually arrange the windows each time the tools are activated.
It would now be interesting to explore how these layouts can be dynamically rearranged, either automatically or interactively, to adapt to the user's change in focus during a data analysis.
Approaches such as automated layout computations, as explored in multi-display VA~\cite{EST19}, could be helpful in this regard.
Moreover, the entire contents of the individual tool windows are displayed unchanged at the moment.
In the future, it might be worthwhile to extract, if possible, only the necessary parts of tool views and display them directly integrated in our unified UI.
Similar to WinCuts~\cite{TMC04} and Fa\c{c}ades~\cite{SCPR06}, this could reduce clutter on the screen, e.g., by removing redundant interface components, while our visualization of the workflow and coordination graph helps to understand the graphical composition.

With respect to the coordination graph, we have so far only considered tool activation and data exchange between tools.
However, during our testing, we also observed the need to synchronize the tools on the parameter level.
These include, in particular, visualization-specific parameters such as applied filters or highlighting as well as selected color scales.
Consistent adjustment of these parameters across tools would result in matched visual output, making composite representations easier to understand on screen.
Addressing all these topics in the future will require additional user studies to assess the applicability of our approach and the potential benefits of the extensions.

\section*{Acknowledgments}
This work has been supported by the German Research Foundation (project UniVA, grant number 380014305).

\printbibliography

\end{document}